\documentclass[12pt,english]{revtex4}
\usepackage{mathptmx}

\usepackage{courier}

\usepackage[T1]{fontenc}
\usepackage[latin9]{inputenc}
\usepackage{amssymb}
\usepackage{esint}

\makeatletter
\@ifundefined{textcolor}{}
{%
 \definecolor{BLACK}{gray}{0}
 \definecolor{WHITE}{gray}{1}
 \definecolor{RED}{rgb}{1,0,0}
 \definecolor{GREEN}{rgb}{0,1,0}
 \definecolor{BLUE}{rgb}{0,0,1}
 \definecolor{CYAN}{cmyk}{1,0,0,0}
 \definecolor{MAGENTA}{cmyk}{0,1,0,0}
 \definecolor{YELLOW}{cmyk}{0,0,1,0}
}

\def\aap{{Astronomy and Astrophys.}}

\def\jgr{{J.~Geophys.~Res.}}
\def\mnras{{MNRAS}}

\def\apj{{\it Astrophys. J.\ }}
		
\def\apjl{{\apj\ \it Lett.\ }}

\def\jgr{{\it J.~Geophys.~Res.\ }}

\def\mnras{{\it Mon. Not. R. Astron. Soc.\ }}
	
\def\ssr{{\it Space Sci. Rev.\ }}

\def\memsai{{\it Memorie della Societa Astronomica Italiana }}

\usepackage{hyperref}

\makeatother

\usepackage{babel}
\begin{document}

\title{Cosmic Ray Confinement and Transport Models for Probing their Putative
Sources}

\author{M.A. Malkov}

\affiliation{CASS and Department of Physics, University of California, San Diego,
La Jolla, CA 92093-0424}
\begin{abstract}
Recent efforts in cosmic ray (CR) confinement and transport theory
are discussed. Three problems are addressed as being crucial for understanding
the present day observations and their possible telltale signs of
the CR origin. The first problem concerns CR behavior right after
their release from a source, such as a supernova remnant (SNR). At
this phase the CRs are confined near the source by self-emitted Alfven
waves. The second is the problem of diffusive propagation of CRs through
the turbulent ISM. This is a seemingly straightforward and long-resolved
problem, but it remains controversial and reveals paradoxes. A resolution
based on the Chapman-Enskog asymptotic CR transport analysis, that
also includes magnetic focusing, is suggested. The third problem is
about a puzzling sharp ($\sim10^{\circ}$) anisotropies in the CR
arrival directions that might bear on important clues of their transport
between the source and observer. The overarching goal is to improve
our understanding of all aspects of the CR's source escape and ensuing
propagation through the galaxy to the level at which their sources
can be identified observationally.
\end{abstract}
\maketitle

\section{Introduction}

Cosmic rays (CR) have been discovered more than a century ago but
the problem of their origin is still with us today. The fundamental
obstacle to identification of their possible sources, such as the
supernova remnant (SNR) shocks, is the CR 'black-box' propagation
through the chaotic magnetic field of the galaxy. With a possible
exception for the highest energy CR ($\gtrsim10^{19}$eV, whose origin
is almost certainly extragalactic \cite{Berezinsky08,BlandfordCRorig14})
most of the CRs arrive from random directions saying nothing about
the locale of their sources. The more surprising is a sharp ($\sim10^{\circ}$)
CR anisotropy discovered by Milagro \cite{Milagro08PRL} with interesting
ramifications due to IceCube, ARGO-YBJ \cite{ARGO13}, HAWC \cite{HAWC_Anis14}
and some other instruments. Had it been created in the source, it
would have been completely erased en route to the Earth. Therefore,
this CR feature is likely to be an imprint of their interaction with
the ISM (interstellar medium) or its local environment (heliosphere
and surroundings). We will discuss this later. More natural is to
start with the CR transport right at their birth place.

SNRs are widely regarded as the most probable source of the bulk of
the CRs \cite{BlandfordCRorig14}. The \emph{joint analysis} of the
broad band observations of the SNRs and CR background spectra at the
Earth should provide the ultimate evidence for this hypothesis. The
analysis faces multiple problems. First, the accelerated CRs manifest
themselves in SNRs only in form of secondary emission, which is usually
difficult to interpret. For example, the super TeV-photons, carefully
counted by the atmospheric Cerenkov telescopes, to testify for the
accelerated protons colliding with the ambient gas \cite{AharDV94,BykovMC00},
can easily be confused with the inverse Compton (IC) photons up-scattered
by accelerated electrons. If this is the case, not much weight can
be added to the argument for the CRs origin in the SNRs, as electrons
comprise only a small fraction ($\sim1-2$\%) of the CR spectrum.
Similarly, in the GeV energy band the emission may come from electron
Bremsstrahlung. The key in both cases, however, is a dense gas in
the SNR surroundings, often present in form of adjacent molecular
clouds (MC). They provide a target for the $pp$ reactions with accelerated
protons. Photons, produced in this reaction should thus come from
MCs, illuminated by the CR protons that have, in turn, escaped from
the source. By contrast, the IC electron emission should come from
the entire volume they fill, as the low-energy photons (such as CMB)
are present everywhere. To use this simple but powerful diagnostic
tool for identification of the source of emission detected by modern
ground based instruments and space observatories (such as the Cerenkov
telescopes HESS, VERITAS, MAGIC and Fermi-LAT, PAMELA, Agile spacecraft
observatories \cite{Aharonian08RPP,TavaniIC443_10,VeritasCasA10,Abdo10W44full,Adriani11,AgileW44_11,MagicW51_12}),
we must have an accurate understanding of the CR propagation from
their sources to the adjacent MCs. 

The next problem is the subsequent interaction of the CRs with the
MC, as their confinement inside the cloud is generally deteriorated
due to the collisional damping of the Alfven waves, that otherwise
would prevent CRs from spreading further rapidly. In addition, this
interaction reveals important clues as to how the spectrum of CRs,
illuminating the MC is different from that in the source, most importantly,
in form of spectral breaks. This aspect of the CR interaction with
MC and their visibility in the gamma-ray band has been discussed in
one of the earlier APS Plasma Physics meetings \cite{MDS_APS12}.
Here we will focus on the ensuing propagation of the CR to the Earth
and their spectral features that they can acquire both in the source
and on the way to us.

Measurements of the CR background spectrum have also advanced significantly.
Much progress has been made in isolating different elements in it.
One of the most striking results was the $\approx0.1$ difference
between the rigidity (momentum to charge ratio) spectral indices of
protons and ${\rm He^{2+}}$ ions. Such deviations have been apparent
for some time, e.g., \cite{ATIC06} but the Pamela spacecraft observatory
measured it with a three-digit accuracy in the 100 GeV energy band
\cite{Adriani11}, which posed a strong challenge to the CR acceleration
and propagation models. Indeed, the ultrarelativistic parts of the
rigidity spectra must be identical, if protons and He$^{2+}$ ions
are accelerated and transported via electromagnetic interactions under
identical conditions. Since He$^{2+}$ has a $0.1$-flatter spectrum,
the difference may be due to its spallation, biased for lower energies
\cite{BlasiChemComp12}. However, such scenario would probably require
stretching the model parameters too much \cite{VladimirMoskPamela11}.
Other interesting possibilities discussed in the literature include
the contribution from multiple SNRs of different types with somewhat
different CR spectra \cite{Bier95,Zatsepin06,Adriani11,Liu15} and
variable $p$/He$^{2+}$ mix along the shock path \cite{DruryEscape11,Ohira11}.
A 'plasma physics' solution that targets the nonrelativistic phase
of acceleration of both species, where the argument of equal rigidity
spectra is irrelevant, was suggested in Ref.\cite{MDSPamela12}. This
explanation is advantageous according to \emph{Occam's razor}, as
it relies on the collisionless shock \emph{intrinsic }properties and
does not require any of the above special conditions.

Apart from the elemental composition, other spectral signatures, such
as the spectral hardening above $E\sim200$ GeV, have been studied
\cite{Adriani11,BlasiChemComp12,Ptuskin13} and provided important
clues for the energy dependent CR transport. Stochasticity of CR sources
and inhomogeneity of transport through the galaxy are now also included
in the models \cite{Tomassetti12ApJ}. These are important for understanding
the large scale CR anisotropy. The most puzzling aspect of the anisotropy
in the CR arrival directions is, in my view, the sharp anisotropy
or the so-called Milagro 'hot spots' which I address later in this
brief review. 

The remainder of the paper is organized as follows. In Sec.\ref{sec:Self-Confinement-of-CRs}
the confinement of CRs released from the source is addressed. In Sec.\ref{sec:Diffusive-and-Hyperdiffusive}
an equation describing diffusive propagation including a hyperdiffusive
term is presented and its relation to the so-called ``telegrapher''
term in the CR transport equation is clarified. In Sec.\ref{sec:Small-Scale-CR-Anisotropy}
possible mechanisms for building a sharp CR anisotropy during their
propagation from the source are discussed.

\section{Self-Confinement of CRs around SNRs\label{sec:Self-Confinement-of-CRs}}

It is widely believed that CRs are accelerated in SNR shocks by the
diffusive mechanism (DSA). The backbone of the DSA is a self-confinement
of accelerated particles supported by their scattering off magnetic
irregularities that particles drive by themselves while streaming
ahead of the shock. Logically, this process should also control the
ensuing propagation (escape) of CRs, at least until their density
drops below the wave instability threshold. At the same time, no consensus
has been reached so far as to how CRs escape the accelerator. The
dividing lines seem to run across the following issues: (i) does the
escape occur isotropically or along the local magnetic field? (ii)
does the scattering by the background MHD turbulence control the CR
propagation alone or self-excited waves need to be included? (iii)
are CRs, that escape SNR, peaked at the highest energy or lower energy
CRs escape as well? \cite{AharAt96,GabiciAharEsc09,Gabici11}\cite{Rosner96}
\cite{GabiciAnisEsc12}\cite{DruryEscape11}\cite{PtuskinPropRev12}.

Adhering to the self-confinement idea, we consider the model that
explicitly includes the self-excited waves. Moreover, in the regions
where magnetic perturbations are weak, i.e. $\delta B^{2}/B^{2}\ll1$,
a field aligned CR transport dominates, as the perpendicular diffusion
is suppressed, $\kappa_{\perp}\simeq\left(\delta B/B\right)^{2}\kappa_{B}\ll\kappa_{\parallel}\simeq\kappa_{B}\left(\delta B/B\right)^{-2}$.
Here $\kappa_{B}$ is the Bohm diffusion coefficient $\kappa_{B}=cr_{g}/3$
with $r_{g}$ being the particle gyroradius. Taking into account the
condition $\left(\delta B^{2}/B^{2}\right)_{{\rm ISM}}\ll1$, such
regime is inevitable outside the source where $\delta B/B\lesssim1$,
as well as at later times of CR propagation, when they are spread
over a large volume and the waves are driven weakly. Moreover, the
self-confinement of CRs propagating away from the accelerator, as
described below, is the continuation of physically the same process
long believed to be at work inside the accelerator, as first suggested
by Bell \cite{Bell78}. From a mathematical standpoint our treatment
below generalizes Bell's steady state solution, obtained in the shock
frame, to the time dependent solution for the CR cloud expanding further
out. This being said, we use the following equations that describe
the CR diffusion and wave generation self-consistently \cite{MetalEsc13}

\begin{equation}
\frac{\partial}{\partial t}P_{{\rm {\rm CR}}}\left(p\right)=\frac{\partial}{\partial z}\frac{\kappa_{{\rm B}}}{I}\frac{\partial P_{{\rm CR}}}{\partial z}\label{eq:dPdt}
\end{equation}

\begin{equation}
\frac{\partial}{\partial t}I=-C_{{\rm A}}\frac{\partial P_{{\rm CR}}}{\partial z}-\Gamma I.\label{eq:dIdt}
\end{equation}
where $C_{A}$ is the Alfv\'en velocity. The dimensionless CR partial
pressure $P_{{\rm CR}}$ is used instead of their distribution function
$f\left(p,t\right)$:

\begin{equation}
P_{{\rm CR}}=\frac{4\pi}{3}\frac{2}{\rho C_{{\rm A}}^{2}}vp^{4}f,\label{eq:PcrDef}
\end{equation}
where $v$ and $p$ are the CR speed and momentum, and $\rho$- the
plasma density. The total CR pressure is normalized to $d\ln p$,
similarly to the wave energy density $I$: 

\[
\frac{\left\langle \delta B^{2}\right\rangle }{8\pi}=\frac{B_{0}^{2}}{8\pi}\int I\left(k\right)d\ln k=\frac{B_{0}^{2}}{8\pi}\int I\left(p\right)d\ln p
\]
The last relation implies a simplified wave-particle resonance condition,
$kr_{g}\left(p\right)=const\sim1$. Most of the works on CR self-confinement
(see \cite{BykBrandMalk13} for a review) use equations largely similar
to eqs.(\ref{eq:dPdt}-\ref{eq:dIdt}) but different assumptions are
made regarding geometry of particle escape from the source, the character
and strength of the wave damping $\Gamma$, and the role of quasilinear
wave saturation. A reasonable choice of the damping mechanism is the
Goldreich-Shridhar (GS) MHD cascade \cite{goldr97}, which seems to
be appropriate in $I\lesssim1$ regime \cite{FarmerGoldr04,BeresnLaz08,YanLazarianEscape12}.
The damping rate in this case is $\Gamma=C_{A}\sqrt{k/l}$, where
$l$ is the outer scale of turbulence, which may be as large as $100pc$
(see, however, Sec.\ref{sec:Small-Scale-CR-Anisotropy}). As $\Gamma$
does not depend on $I$ and can be considered also as coordinate independent,
it allows the following ('quasilinear') integral of the system given
by eqs.(\ref{eq:dPdt}) and (\ref{eq:dIdt}):

\begin{equation}
P_{{\rm CR}}\left(z,t\right)=P_{{\rm CR}0}\left(z\right)-\frac{\kappa_{{\rm B}}}{C_{{\rm A}}}\frac{\partial}{\partial z}\ln\frac{I\left(z,t\right)}{I_{0}\left(z\right)}\label{eq:QLint}
\end{equation}
Here $P_{{\rm CR}0}\left(z\right)$ and $I_{0}\left(z\right)$ are
the initial distributions of the CR partial pressure and the wave
energy density (see \cite{MetalEsc13} for more general treatment).
Substituting $P_{CR}$ in eq.(\ref{eq:dIdt}) we arrive at the following
diffusion equation for $I$
\[
\frac{\partial I}{\partial t}=\frac{\partial}{\partial z}\frac{\kappa_{B}}{I}\frac{\partial I}{\partial z}-\Gamma I-C_{A}\frac{\partial P_{CR0}}{\partial z}
\]
Outside of the region where $P_{CR0}\neq0,$ the last term on the
r.h.s. is absent, while the second term may be eliminated by replacing
$I\exp\left(\Gamma t\right)\to I$, $\intop_{0}^{t}\exp\left(\Gamma t\right)dt\to t$.
However, if $\Gamma$ is taken in a GS-form, it is fairly small due
to the factor $\sqrt{r_{g}/l}\ll1$. We may simply neglect it. The
solution for $I$ and $P_{CR}\left(z,t\right)$ may be found in an
implicit form (see \cite{MetalEsc13} for details). However, there
exists a very accurate convenient interpolation formula that can be
represented as follows

\begin{equation}
P_{{\rm CR}}=\frac{2\kappa_{{\rm B}}\left(p\right)}{C_{{\rm A}}^{3/2}\sqrt{at}}\left[\zeta^{5/3}+\left(D_{{\rm NL}}\right)^{5/6}\right]^{-3/5}e^{-\zeta^{2}/4D_{{\rm ISM}}}\label{eq:Ufit}
\end{equation}
where $a$ is the size of the initial CR cloud, $\zeta=z/\sqrt{C_{A}at}$,
$D_{{\rm NL}}=F\left(\Pi\right)\cdot D_{{\rm ISM}}\exp\left(-\Pi\right)$,
with $\Pi$ being a normalized integrated pressure

\[
\Pi=\frac{C_{A}}{\kappa_{B}}\intop_{0}^{\infty}P_{CR}dz
\]
The function $F$, behaves as follows $F\left(\Pi\right)\simeq2e\approx5.4$,
for $\Pi\gg1$ and $F\left(\Pi\right)\simeq2\pi\Pi^{-2}$, for $\Pi\ll1$.
Here $D_{{\rm ISM}}$ is a normalized background diffusivity $D_{{\rm ISM}}=\kappa_{B}/aC_{A}I_{{\rm ISM}}$. 

To summarize these results, the self-regulated CR escape from a source
is characterized by their distribution (partial pressure) comprising
the following three zones: (i) a quasi-plateau (core) at small $z/\sqrt{t}<\sqrt{D_{{\rm NL}}}$
of the height $\sim\left(D_{{\rm NL}}t\right)^{-1/2}$. It is elevated
by a factor $\sim\Pi^{-1}\exp\left(\Pi/2\right)\gg1$, compared to
the test particle solution because of the strong quasi-linear suppression
of the CR diffusion coefficient with respect to its background (test
particle) value $D_{{\rm ISM}}$: $D_{{\rm NL}}\sim D_{{\rm ISM}}\exp\left(-\Pi\right)$
(ii) next to the core, where $\sqrt{D_{{\rm NL}}}<z/\sqrt{t}<\sqrt{D_{{\rm ISM}}}$,
the profile is scale invariant, $P_{{\rm CR}}\propto1/z$. The CR
distribution in this ``pedestal'' region is fully self-regulated
and independent of $\Pi$ and $D_{{\rm ISM}}$ for $\Pi\gg1$, (iii)
the tail of the distribution at $z/\sqrt{t}>\sqrt{D_{{\rm ISM}}}$
is similar \emph{in shape }to the test particle solution in 1D but
it saturates with $\Pi\gg1$, so that the CR partial pressure is $\propto\left(D_{{\rm ISM}}t\right)^{-1/2}\exp\left(-z^{2}/4D_{{\rm ISM}}t\right)$,
\emph{independent of the strength of the CR source} $\Pi$, in contrast
to the test-particle regime in which it scales as $\propto\Pi$, ($\Pi\lesssim1$).
Because of the CR diffusivity reduction, the half-life of the CR cloud
is increased and its width is decreased, compared to the test particle
solution. Depending on the functions $\Pi\left(p\right)$ and $D_{{\rm ISM}}\left(p\right)$,
the resulting CR spectrum generally develops a spectral break for
the fixed values of $z$ and $t$ at the CR momentum $p$ determined
by the following relation: $z^{2}/t\sim D_{{\rm NL}}\left(p\right)\sim D_{{\rm ISM}}\exp\left(-\Pi\right)$.

\section{Diffusive and Hyperdiffusive CR Transport with magnetic focusing\label{sec:Diffusive-and-Hyperdiffusive}}

Propagating away from their sources, CRs are pitch-angle scattered
on weak ISM magnetic irregularities. A seemingly straightforward reduction
of kinetic CR description to their spatial transport leads to a diffusive
approximation which has the well-known defect of causality violation.
There have been attempts at an alternative approach based on the ``telegrapher''
equation. However, its derivations often lack rigor and transparency
and had not been performed to the required (as we show below, fourth
order) accuracy. The problem can be formulated very plainly: How to
describe CR transport by only their isotropic component, when the
anisotropic one is suppressed by the frequent scattering? 

The angular distribution of CRs is described by the function $f\left(\mu,t,z\right)$
\cite{VVSQL62,Jokipii66}:

\begin{equation}
\frac{\partial f}{\partial t}-\frac{\partial}{\partial\mu}D\left(\mu\right)\left(1-\mu^{2}\right)\frac{\partial f}{\partial\mu}=-\varepsilon\left(\mu\frac{\partial f}{\partial z}+\frac{\sigma}{2}\left(1-\mu^{2}\right)\frac{\partial f}{\partial\mu}\right)\label{eq:BoltzmannRescaled}
\end{equation}
Here $\varepsilon=v/l\nu$ is the small parameter of the problem,
with $v$ being the particle velocity, $l$- characteristic scale
and $\nu$- scattering frequency. The dimensionless magnetic mirror
inverse scale $\sigma=-B^{-1}\partial B/\partial z$, $z$ points
in the local field direction and is measured in the units of $l$,
time - in $\nu^{-1}$, $\mu$ is the cosine of the pitch angle, while
$D\left(\mu\right)\sim1$ depends on the spectrum of magnetic fluctuations.
The isotropic reduction scheme requires a multi-time asymptotic (Chapman-Enskog)
expansion. So, we introduce a set of formally independent time variables
$t\to t_{0},\; t_{1},\dots$, so that 

\begin{equation}
\frac{\partial}{\partial t}=\frac{\partial}{\partial t_{0}}+\varepsilon\frac{\partial}{\partial t_{1}}+\varepsilon^{2}\frac{\partial}{\partial t_{2}}\dots\label{eq:TimeDerExpan}
\end{equation}
which leads to the following hierarchy of equations 

\begin{eqnarray}
\frac{\partial f_{n}}{\partial t_{0}}-\frac{\partial}{\partial\mu}D\left(\mu\right)\left(1-\mu^{2}\right)\frac{\partial f_{n}}{\partial\mu} & = & -\mu\frac{\partial f_{n-1}}{\partial z}-\frac{\sigma}{2}\left(1-\mu^{2}\right)\frac{\partial f_{n-1}}{\partial\mu}-\sum_{k=1}^{n}\frac{\partial f_{n-k}}{\partial t_{k}}\label{eq:fnGen}\\
 & \equiv & \mathcal{L}_{n-1}\left[f\right]\left(t_{0},\dots,t_{n};\mu,z\right)\nonumber 
\end{eqnarray}
where $f=f_{0}+\varepsilon f_{1}+\varepsilon^{2}f_{2}+\dots$ and
the conditions $f_{n<0}=0$ are implied. Using the above expansion,
one may obtain an equation for the isotropic part $f_{0}=\left\langle f\right\rangle \equiv\left(1/2\right)\intop fd\mu$
to arbitrary order in $\varepsilon$. By construction, in no order
of approximation will higher time derivatives emerge, as was obviously
devised in the Chapman-Enskog method. We terminate this process at
the fourth order, $\varepsilon^{4}$. This is the lowest approximation
required to clarify the origin of the telegrapher equation. Higher
order terms can in principle be calculated at the expense of involved
algebra but such calculations would be of no avail. So, our result
is as follows \cite{2015arXiv150201799M}

\begin{eqnarray}
\frac{\partial f_{0}}{\partial t} & = & \frac{\varepsilon^{2}}{4}\partial_{z}^{\prime}\left\{ \kappa-\varepsilon\partial_{z}^{\prime\prime}\left\langle \mu W^{2}\right\rangle \right.\nonumber \\
 & - & \left.\frac{\varepsilon^{2}}{2}\left[\left(\partial_{z}^{\prime\prime}\right)^{2}\left\langle W^{2}\left(\kappa-U^{\prime}\right)\right\rangle -\frac{1}{2}\partial_{z}^{\prime}\partial_{z}\left\langle \frac{\left[\kappa\left(1-\mu\right)+U\right]^{2}}{D\left(1-\mu^{2}\right)}\right\rangle \right]\right\} \frac{\partial f_{0}}{\partial z}\label{eq:Master4}
\end{eqnarray}
where $\partial_{z}^{\prime}=\partial_{z}+\sigma$ and $\partial_{z}^{\prime\prime}=\partial_{z}+\sigma/2$.
The coefficients are defined as $\partial W/\partial\mu=1/D,\;\left\langle W\right\rangle =0$,
$U^{\prime}\left(\mu\right)\equiv\partial U/\partial\mu\equiv\left(1-\mu^{2}\right)/D$,
$U\left(-1\right)=0,$ and $\kappa=\left(1/2\right)U\left(1\right)$.

\subsubsection{Producing the telegrapher term}

Within the employed Chapman-Enskog expansion, the equation for $f_{0}$
remains evolutionary in all orders of $\varepsilon$, so no telegrapher
term appears. Such term is usually obtained either without clear ordering,
e.g. \cite{LitvSchlick13}, using specific $D\left(\mu\right)$, e.g.
\cite{SchwadronTelegraph94}, or by truncation of eigenfunction expansion,
where the discarded terms may be of the same order in small parameter
as those retained, e.g., \cite{Earl73,Earl96}. In most of these treatments
care has not been exercised to eliminate the short time scales which
are irrelevant to the long-time evolution of the isotropic part of
the CR distribution, sought by these reduction schemes. Instead, they
retain the second time derivative which changes the type of the resulting
transport equation to hyperbolic. As we show below, the second order
time derivative term can be recovered from the Chapman-Enskog expansion.

To resolve the above controversy, we simplify eq.(\ref{eq:Master4})
by removing terms unimportant for the controversy. First, we may set
$\sigma=0$ and assume the scattering symmetry, $D\left(-\mu\right)=D\left(\mu\right)$,
to remove the term $\sim\varepsilon^{3}$, as the $\partial^{3}/\partial z^{3}$
term is not included in the telegrapher equation derived for magnetic
focusing by, e.g., \cite{LitvSchlick13}. Using these simplifications
and the slow time $T=\varepsilon^{2}t/4$, eq.(\ref{eq:Master4})
rewrites:

\begin{equation}
\frac{\partial f_{0}}{\partial T}=\kappa\frac{\partial^{2}f_{0}}{\partial z^{2}}-\varepsilon^{2}K\frac{\partial^{4}f_{0}}{\partial z^{4}}\label{eq:MasterSimple}
\end{equation}
where $K$ is the hyper-diffusion coefficient 

\begin{equation}
K=\frac{1}{2}\left\langle W^{2}\left(\kappa-U^{\prime}\right)-\frac{1}{2}\frac{\left[\kappa\left(1-\mu\right)+U\right]^{2}}{D\left(1-\mu^{2}\right)}\right\rangle \label{eq:Kdef}
\end{equation}
To the same order in $\varepsilon\ll1$, the last equation can be
rewritten as follows

\begin{equation}
\frac{\partial f_{0}}{\partial T}=\kappa\frac{\partial^{2}f_{0}}{\partial z^{2}}-\tau\frac{\partial^{2}f_{0}}{\partial T^{2}},\label{eq:TelegrDerived}
\end{equation}
where $\tau=\varepsilon^{2}K/\kappa^{2}$. This equation has indeed
the form of a telegrapher equation. However, the comparison of eq.(\ref{eq:MasterSimple})
with, e.g., the telegrapher eq.(15) in Ref.\cite{LitvSchlick13} shows
that, the coefficient $\tau$ in eq.(\ref{eq:TelegrDerived}) is substantially
different. The reason is that the equation of Ref.\cite{LitvSchlick13}
has been obtained by a formal iteration not accounting for all the
fourth order terms, the telegrapher term actually originates from.
Note that \cite{SchwadronTelegraph94} give an expression for $\tau$
which is probably consistent with the result above, but their $\tau$
was obtained for a power-law $D\left(\mu\right)$, so the comparison
would take extra steps, not worth doing in this short review. More
importantly, the telegrapher term in eq.(\ref{eq:TelegrDerived})
has a small parameter (at highest derivative). The role of such terms
is known from the boundary layer problems. They become crucial near
and inside the boundary layer, thus determining its structure and
scale. In the context of the telegrapher equation, the boundary layer
translates into the initial relaxation phase of the CR distribution.
This relaxation is associated with the small scale CR anisotropy in
$f_{n}$ which quickly decays. It should be noted that if a simplified
collision term (BGK, or $\tau$- approximation) is used instead of
the pitch-angle diffusion in eq.(\ref{eq:BoltzmannRescaled}), the
telegrapher equation can be accurately derived with no recourse to
hyperdiffusion \cite{WebbTelegr06}.

To conclude this section, by comparison with the telegrapher equation,
the classic Chapman-Enskog is a considerably more suitable and flexible
tool to describe the long-time CR propagation, although the telegrapher
version (with corrected transport coefficient) may still be useful
for studying the magnetically focused CR transport, e.g. \cite{Effenberger2014}.
Efforts on improving the CR diffusion models, where their drawbacks
are important, need to address the lower level transport, including
anisotropic component of the CR distribution, directly. Recent work
can be found in, e.g., \cite{AloisBerezSuperLum09} and in the next
section. Splitting the particle distribution in scattered and unscattered
categories is another useful approach, e.g., \cite{Webb2000,Zank2000}.
However, when the diffusive treatment is well within the method's
validity range (weakly anisotropic spatially smooth CR distributions)
neither the telegrapher term nor hyperdiffusivity are essential to
the CR transport (see \cite{2015arXiv150201799M} for more discussion).

\section{Small-Scale CR Anisotropy\label{sec:Small-Scale-CR-Anisotropy}}

CR acceleration (e.g. DSA) and propagation models, as discussed in
the two preceding sections, typically predict only a large scale,
dipolar anisotropy. It would emerge as a small $f_{1}\propto\mu$
correction to $f_{0}\gg f_{1}$, produced by localized sources, and
can be easily obtained within the treatment outlined in Sec.\ref{sec:Diffusive-and-Hyperdiffusive}.
The same is true for the CR self-confinement problem considered in
Sec.\ref{sec:Self-Confinement-of-CRs}, if the small anisotropic correction
is taken into account. Now that we expect the CR propagation in the
essentially stochastic magnetic fields to be largely ergodic, there
is no obvious reason for a significantly sharper than the dipolar
anisotropy. Yet observations show that narrow ($\sim10^{\circ}$)
CR beams do exist \cite{Milagro08PRL,ARGO13,HAWC_Anis14}. They shed
new lights on the CR propagation from, and even their acceleration
in, putative sources and need to be understood. 

A number of scenarios have been suggested to explain the tightly collimated
beams. They include magnetic nozzle focusing \cite{DruryAharMILAGRO08},
propagation effects from local SNR \cite{SalvatiMilagro08}, acceleration
in the heliotail \cite{LazarMilagro10,Desiati10} and heliosheath
propagation effects \cite{Desiati13}. Although being plausible in
principle, those explanations impose significant constraints on the
relevant parameters and processes. For example, the magnetic mirror
ratio must be rather strong to produce $\sim10^{\circ}$ anisotropy,
and quite a strong magnetic field in the heliotail is required to
confine 10 TeV protons and make the proposed acceleration mechanism
work efficiently. Conceptually different scenarios \cite{Milagro10,Giacinti12,Ahlers14}
essentially attempt at generating small-scale anisotropy out of the
large-scale one by exploiting aspects of interactions between the
CRs and MHD turbulence in the ISM. At the first glance, precisely
the opposite should occur and the task is clearly of a kind of ``squeezing
blood out of stone''. From a purely mathematical perspective, using
certain properties of the particle propagator, these models produce
multipoles out of the dipolar component over a long distance (up to
a few 100pc) of particle propagation. At this point, however, the
approaches deviate strongly from one another.

In Ref.\cite{Ahlers14}, an interesting technique is employed to generate
higher multipoles from the available dipole by using the Liouville's
theorem. It is not clear, however, whether the introduction of a simplified
collision term in a BGK-form, is justified for the treatment of the
small-scale anisotropy. The preferred collision operator is the differential
one which is much more efficient at smoothing small-scale anisotropies
(see, e.g., previous section). Attacking the same problem from a different
angle, the authors of Ref.\cite{Giacinti12} rightly state that, although
the scattering fields are random, we do not really need to perform
an ensemble average, as the current MHD turbulence is static for the
limited time observations and fast CRs. There are at least two tests
to propose for this explanation. First, as this is actually a magnetic
lensing effect with a very long particle path ($L\gg r_{g}$), small
variations of magnetic configuration may produce significant changes
in arrival directions of narrow beams. Indeed the relevant scale of
the turbulent field is $r_{g}$, so the time scale is $\tau\sim r_{g}/(V_{A}+U_{HS})$
with the Alfven and the heliosphere velocities in denominator. The
median Milagro energy is $\sim1$TeV, so for $V_{A}+U_{HS}\simeq$50km/s
and $B=4\mu G$ one obtains $\tau\lesssim10$yrs. This may be close
enough to the time difference between Milagro and ARGO/HAWC more recent
observations. And yes, changes are being observed but they are not
quite significant and HAWC is not fully operational yet, so more observations
are required and they are underway \cite{HAWC_Anis14}. The second
test has, in fact, already been performed by the authors of Ref.\cite{Giacinti12}.
Since CRs interact with the static magnetic fields, their dynamics
may be regarded as almost ergodic (strong orbit mixing) on every isoenergetic
surface in phase space. Small deviations from ergodicity are responsible
for the hot spots in arrival directions. Moving from one energy surface
to the next by $\Delta E\sim E$ should strongly decorrelate the spots,
since $\Delta r_{g}\sim r_{g}$ for them. This is indeed observed
in simulations carried out in Ref.\cite{Giacinti12}. The upcoming
improvements in the energy spectra measurements \cite{HAWC_Anis14}
should substantiate such tests quantitatively and help to discriminate
between different mechanisms.

The approach of Ref.\cite{Milagro10} is also based on the CR interaction
with the ISM turbulence, but includes ensemble averaging, thus removing
the above concerns with the short time variability (except for the
possible heliospheric variations \cite{2012AIPC.1436..190M}). The
beam direction is assumed to be along the local large scale magnetic
field ($l_{loc}\gg r_{g}$), to minimize the curvature and gradient
drifts, that would otherwise evacuate particles from the magnetic
tube connecting observer with the source, since the drifts increase
with the pitch angle. The following assumptions are made to obtain
the beam collimation: (i)\emph{ large scale} anisotropic distribution
of CRs (generated, for example, by a nearby accelerator, such as a
SNR, magnetically connected with the Earth) and (ii) Goldreich-Shridhar
\cite{Goldr95} (GS) cascade of Alfvenic turbulence originating from
a specific scale $l$, which is the longest scale relevant to the
wave-particle interactions. 

It is found that the CR distribution develops a characteristic angular
shape consisting of a large scale anisotropic part (first eigenfunction
of the pitch-angle scattering operator) superposed by a beam, sharply
focused in the momentum space along the local field. The large scale
anisotropy carries the \emph{momentum dependence }of the source. The
following four quantities are tightly constrained by the turbulence
scale $l$: (1) the beam angular width, that increases with momentum
as $\propto\sqrt{p}$ (2) its fractional excess (with respect to the
large scale anisotropic component), that increases as $\propto p$,
(3) the maximum momentum, beyond which the beam is destroyed via instability,
$p_{{\rm max}}$. If the large scale anisotropy originates from a
nearby source, magnetically connected with the Earth, the model predicts
(4) the range of possible distances to this source, $l_{S}\sim100-200$pc.
If such source is absent, this range corresponds to the beam collimation
length, also a few 100pc, with the large scale anisotropy originating
from the smooth omnigalactic CR gradient. This scale is consistent
with the beam collimation length, obtained numerically in \cite{Giacinti12}.

If the turbulence outer scale $l$ is considered unknown, it can be
inferred from any of the first three quantities (1-3) as measured
by MILAGRO. All the three quantities consistently imply the same scale
$l\simeq$1 pc. The calculated beam maximum momentum encouragingly
agrees with that measured by MILAGRO ($p_{{\rm max}}\sim10$ TeV/c).
The theoretical value for the angular width of the beam is found to
be $\Delta\vartheta\simeq4\sqrt{\epsilon}$, where $\epsilon=r_{g}\left(p\right)/l\ll1$.
The beam fractional excess related to the large scale anisotropic
part of the CR distribution is $\simeq50\epsilon$. Both quantities
also match the Milagro results near its median energy, that is for
$E\sim1-2$ TeV. So, the beam has a momentum scaling that is one power
shallower than the CR carrier, it is drawn from. One interesting conjecture
from the $l\simeq$1 pc requirement is that the proton 'knee' at $\simeq3$
PeV and the beam are of the same origin, as these particles may provide
the required outer scale for the MHD turbulence, $r_{g}\sim l$. Another
possibility is to employ the spiral-arm 1-pc value for $l$, as suggested
in \cite{OuterScale08}. 
\begin{acknowledgments}
I'm indebted to F. Aharonian, P. Diamond, L. Drury, I. Moskalenko
and R. Sagdeev for collaboration on the topics discussed in this paper.
Suggestions by the anonymous referee are also greatly appreciated.
This work was supported by the NASA ATP-program under the Grant NNX14AH36G.
\end{acknowledgments}

\end{document}